\documentclass[12pt]{article}

\usepackage[english]{babel}

\usepackage[letterpaper,top=2cm,bottom=2cm,left=3cm,right=3cm,marginparwidth=1.75cm]{geometry}

\usepackage{amsmath}
\usepackage{amsfonts}
\usepackage{graphicx}
\usepackage{subcaption}
\usepackage[colorlinks=true, allcolors=blue]{hyperref}
\usepackage{lipsum}
\usepackage{tikz}
\usetikzlibrary{calc}
\usepackage[misc]{ifsym}

\newcommand\blfootnote[1]{%
  \begingroup
  \renewcommand\thefootnote{}\footnote{#1}%
  \addtocounter{footnote}{-1}%
  \endgroup
}

\author{You}
\title{Soumission de projet de numéro thématique}

\begin{document}

\setcounter{section}{3}
\section*{Chapter 3 \\ A nonlinear dead-time compensation method for path tracking control}
\textbf{Karin Festl, Michael Stolz}\blfootnote{\hspace{-6mm}K. Festl (\Letter) $\cdot$ M. Stolz\\
Virtual Vehicle Research GmbH, Graz, Austria\\
e-mail: karin.festl@v2c2.at}

\textbf{Abstract:} In the realm of autonomous vehicle technologies and advanced driver assistance systems, precise and reliable path tracking controllers are vital for safe and efficient navigation. However the presence of dead time in the vehicle control systems poses a challenge to real-world systems. Input and output delays are caused by factors like sensor processing and mechanical response and can range up to a few hundred milliseconds. This chapter addresses the problem of dead time in path tracking control and proposes a method to compensate the dead time.
The proposed solution involves a nonlinear prediction model, in a structure similar to the Smith predictor, but incorporating the kinematic behavior of the vehicle plant system. The implementation avoids numeric integration or optimization, enabling a fast execution.
Simulation tests with various controllers and disturbances, including dead-time uncertainty, demonstrate the efficacy of the dead-time compensation method. Results indicate improved control performance in all tested scenarios.\par
\textbf{Keywords:} dead-time compensator, path tracking, nonlinear control

\subsection{Introduction}

In the area of autonomous vehicle technologies and advanced driver assistance systems, the precision and reliability of path tracking controllers play a pivotal role in ensuring safe and efficient navigation. However, real-world systems often face challenges, one of which is the presence of dead time. The dead time composes of both a delay between the output of a steering command and its actual execution (output delay) and of the time that passes until a state feedback value is generated from the actual vehicle state (input delay). Dead time can result from various factors, including sensor processing, communication latency, and mechanical response.

Addressing dead time is crucial for achieving accurate and robust vehicle control. For many advanced control strategies, as for example optimal predictive control, dead-time compensation is intrinsic. However, in dynamic environments where rapid decision-making and execution is essential, faster control algorithms are used, which often require additional modifications to compensate dead time.
The Smith predictor is a well known method for compensating dead time. In its simplest form, it is not applicable to unstable plants or plants with an integrator. With modifications such as the Smith-Åström predictor \cite{hoelzl}, the dead-time compensation can be applied on a broad range of control systems. Similar approaches are also applied to the path tracking control problem: In \cite{nonholonomic_outputdelay} an observer similar to Luenberger observer is applied to the nonlinear plant system. In \cite{smith_pred_cntrl}, an optimal predictor and a Smith predictor are applied to the linearized vehicle model plant, comparing their robustness. 

In this chapter, we propose a nonlinear method for dead-time compensation that modifies the state feedback similar to the Smith predictor, however, by incorporating the kinematic behavior of the path tracking plant system, it is possible to apply the predictive correction directly to the unstable nonlinear plant system. 
While this method is based on model state prediction, it does not involve numeric integration or optimization but can be implemented in explicit form. We test the dead-time compensation in simulation with different controllers and in presence of different kinds of disturbances and uncertainties, including dead-time uncertainty. The results show that the dead-time compensation improves the control performance in all tested scenarios.

The chapter starts by defining the dead-time problem and the desired behavior of the system with dead-time compensation. After that, the solution is derived and demonstrated in a short example.

\subsection{Problem definition}
Suppose there is a path tracking controller that shows the desired performance for a given use case, when there is no dead time in the system. When dead time emerges in the system, the controller adaptions are required. We first describe the control system comprising dead time and afterwards define the desired properties of a dead-time compensation method.

\subsubsection{Path tracking with dead time}
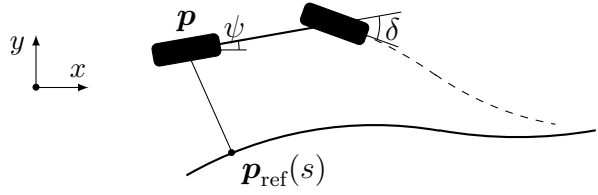
\begin{figure}[tb] 
	\centering 
	\begin{tikzpicture}[]
 \def\vehicle[#1]#2(#3)#4(#5,#6,#7,#8)(#9)
  {\node [draw, #1, shape=rectangle, minimum width=#8*0.3cm, minimum height=#8*0.1cm,rotate=#6, rounded corners=2pt,fill, fill opacity=0.5] (#9rear) at (#3) {};
\node [draw, #1, shape=rectangle, minimum width=#8*0.3cm, minimum height=#8*0.1cm,rotate=#6+#7, rounded corners=2pt,fill, fill opacity=0.5](#9front)at ($(#9rear)+(#6:#5)$) {};
\draw[thick,#1] (#9front.center)--(#9rear.center)coordinate[pos=0](#9cg){};}

 \def\vehiclelead[#1]#2(#3)#4(#5,#6,#7,#8)(#9)
  {\node [#1, shape=rectangle, minimum width=#8*0.3cm, minimum height=#8*0.1cm,rotate=#6, rounded corners=2pt, fill opacity=0.5] (#9rear) at (#3) {};
\node [draw, #1, shape=rectangle, minimum width=#8*0.3cm, minimum height=#8*0.1cm,rotate=#6+#7, rounded corners=2pt,fill, fill opacity=0.5](#9front)at ($(#9rear)+(#6:#5)$) {};
\draw[#1] (#9front.center)--(#9rear.center)coordinate[pos=0](#9cg){};}

 \def\arclabel[#1]#2(#3)#4(#5,#6,#7,#8)(#9)
 {\draw[#1](#3)--++(#5:#7);
\draw[#1](#3)--++(#6:#7);
\draw (#3) ++(#5:#8) arc (#5:#6:#8)node[pos=0.5,xshift=-0.2cm,yshift=-0.1cm]{#9};}

\def\centerarc[#1](#2)(#3:#4:#5)
    { \draw[#1] ($(#2)+({#5*cos(#3)},{#5*sin(#3)})$) arc (#3:#4:#5); }

\def\lengthlabel[#1](#2)(#3)(#4:#5)(#6)
    {\draw (#2)--++(#4+90:#5+0.1);
     \draw (#3)--++(#4+90:#5+0.1);
    \draw (#2)--++(#4+90:#5-0.1);
     \draw (#3)--++(#4+90:#5-0.1);
    \draw (#2)--++(#4+90:#5)coordinate(l11);
     \draw (#3)--++(#4+90:#5)coordinate(l12);
    \draw [latex-latex](l11)--(l12)node[anchor=south, pos=0.5,rotate=#4]{#6};
 }

\def\markdot(#1) 
{
\draw ($(#1)-(0,0.02)$) node{\Huge$\cdot$};
}


\def\psiveh{10}
\def\deltaveh{-30}
\vehicle[](-0.5,0.5)(2,\psiveh,\deltaveh,3)(veh);

\markdot(vehrear.center)
\draw(vehrear.center)node[anchor=south,yshift=3]{$\boldsymbol{p}$};
\markdot(vehfront.center)

\arclabel[](vehrear.center)(0,\psiveh,0.8,0.7)();
\draw (vehrear.center) node[anchor=west,yshift=8,xshift=10]{$\psi$};

\arclabel[](vehfront.center)(\psiveh,\psiveh+\deltaveh,0.9,0.6)();
\draw (vehfront.center) node[anchor=west,yshift=-2,xshift=15]{$\delta$};



 \def\radr{3.45}
\path(vehrear.center)--++(70+90:\radr)coordinate(cent);
\centerarc[thick](2,-5.5)(80:120:5);
\centerarc[thick](3.9,5.34)(-80:-100:6);

\centerarc[dashed](-0.1,-3.9)(70:54:5);
\centerarc[dashed](5.1,3.45)(-125:-100:4);

\draw [-latex] (-2.5,0) -- (-2.5,.7) node[anchor=east,pos=0.8]{\small$y$};
\draw [-latex] (-2.5,0) -- (-1.8,0) node[anchor=south,pos=0.8]{\small$x$};
\draw (-2.5,-0.02) node[]{\Huge$\cdot$};

\draw[](vehrear.center)--(0.1,-0.85);
\node at (0.8,-1.1) {$\boldsymbol{p}_\text{ref}(s)$};
\markdot(0.1,-0.87)

\end{tikzpicture} 
	\caption{Vehicle with reference point $\boldsymbol{p}$ in the rear axle approaching a reference path $\boldsymbol{p}_\text{ref}(s)$.}
	\label{fig:problem} 
\end{figure}

The path tracking (or path stabilization) problem can be described as controlling the steering of a vehicle such that it approaches and follows a reference or target path. More formally, we need to find a feedback law $\delta\left(\boldsymbol{p},\boldsymbol{p}_\text{ref}\right)$ such that
\begin{equation}\label{eqn:cntrl_goal}
	\exists s: \lim_{t\to\infty}||\boldsymbol{p}(t) - \boldsymbol{p}_\text{ref}(s)|| \leq \varepsilon
\end{equation}
Where $\boldsymbol{p}(t)$ is the vehicle reference point at time $t$, $\boldsymbol{p}_\text{ref}(s)$ is the reference path (we define the path as parametric curve with s being the length along the curve) and $\varepsilon$ is a tolerance distance. In other words, we require a feedback law that steers the minimum distance between vehicle and reference path to a vicinity around $0$. Fig.~\ref{fig:problem} illustrates the vehicle with reference point $\boldsymbol{p}$ and the reference path $\boldsymbol{p}_\text{ref}(s)$.

In the simplest case, the vehicle dynamics $d/dt\, \boldsymbol{p}(t) = \dot{\boldsymbol{p}}(t)$ can be modelled by a kinematic single track model:
\begin{subequations}\label{eqn:kin0}
	\begin{align}
		\boldsymbol{x} &= \begin{bmatrix}
			p_x & p_y & \psi
		\end{bmatrix}^T = \begin{bmatrix}
		\boldsymbol{p} & \psi
		\end{bmatrix}^T\\
	\boldsymbol{\dot x} &= 
	 \begin{bmatrix}
	\cos\psi \\ \sin\psi \\ \tan(\delta)/l
\end{bmatrix}\cdot v
	\end{align}
\end{subequations}

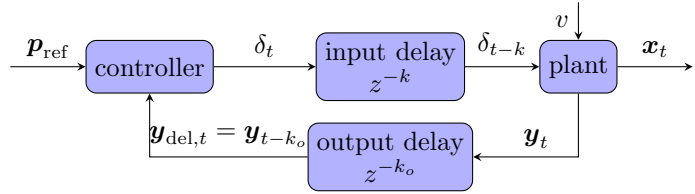
\begin{figure}[tb] 
	\centering 
	\footnotesize
\begin{tikzpicture}

\def\blockwidth{1.5cm}
\def\blockheight{0.7cm}


\node[draw,
rounded corners,
	minimum width = \blockwidth,
    minimum height=\blockheight,
    fill=blue!30,
align=center
] (cntrl) at (0,0){controller};

\node[draw,
rounded corners,
	minimum width = \blockwidth*0.5,
    minimum height=\blockheight,
    fill=blue!30,
	align=center
] (p1) at ($(cntrl)+(3.2,0)$){input delay\\$z^{-k}$};

\node[draw,
rounded corners,
	minimum width = \blockwidth*0.5,
    minimum height=\blockheight,
    fill=blue!30,
	align=left
] (p2) at ($(p1)+(2.5,0)$){plant};

\node[draw,
rounded corners,
	minimum width = \blockwidth*0.5,
    minimum height=\blockheight,
    fill=blue!30,
	align=center
] (p3) at ($(p1)+(0,-1.2)$){output delay\\$z^{-k_o}$};




\draw[-stealth] (cntrl.east)--(p1.west) node[pos=0.5,above]{$\delta_t$};
\draw[-stealth] (p1.east)--(p2.west) node[pos=0.5,above]{$\delta_{t-k}$};
\draw[-stealth](p2.east)--++(1,0)node[pos=0.5,above]{$\boldsymbol{x}_t$};
\draw[stealth-](p3.east)-|(p2.south)node[pos=0.3,above]{$\boldsymbol{y}_t$};
\draw[stealth-](cntrl.south)|-(p3.west)node[pos=0.75,above]{$\boldsymbol{y}_{\text{del},t} = \boldsymbol{y}_{t-k_o}$};

\draw[stealth-](cntrl.west)--++(-1,0)node[pos=0.5,above]{$\boldsymbol{p}_\text{ref}$};

\draw[stealth-](p2.north)--++(0,0.5)node[pos=0.5,left]{$v$};

\end{tikzpicture} 
	\caption{Structure of the controlled plant with input and output delay.}
	\label{fig:cntrl_plant} 
\end{figure}

Where $\delta$ is the steering angle, $v$ is the velocity and $l$ is the wheelbase of the vehicle.
The real vehicle behavior will differ from this model due to several effects which may or may not be considered by the controller. Regardless, the dead-time compensation method will be based on this model.
Moreover, we will formulate the approaches in discrete time: we denote with $\xi_i$ any variable $\xi$ at time $t=i\cdot \Delta t$, where $\Delta t$ is the constant sample time.
The dynamic equations of the kinematic single track model in presence of input delay $\Delta t_\text{I}$ and output delay $\Delta t_\text{O}$ is:
\begin{subequations}\label{eqn:kin_del}
	\begin{align}
	\boldsymbol{x}_{\text{del},i+1} &= \boldsymbol{x}_{\text{del},i} + \int_{i\cdot\Delta t}^{(i+1)\cdot \Delta t}
	 \begin{bmatrix}
	\cos\varphi(\tau) \\ \sin\varphi(\tau) \\ \tan(\delta_{i-k_\text{I}})/l
\end{bmatrix}\cdot v_i \cdot d\tau \\
\boldsymbol{y}_{\text{del},i} &= \boldsymbol{y}_{i-k_\text{O}} = \boldsymbol{x}_{\text{del},i} \cdot q^{-k_\text{O}}
\end{align}
\end{subequations}
Where $\boldsymbol{y}_\text{del}$ is the delayed state feedback, $k_\text{I} = \Delta t_\text{I} / \Delta t$, $k_\text{O} = \Delta t_\text{O} / \Delta t$ are the discrete delays and $q$ is the forward shift operator as defined in \cite{hanus}: $q\cdot \xi_i = \xi_{i+1}$.
In Fig.~\ref{fig:cntrl_plant}, the structure of the controlled planned in the presence of input and output delay is shown.

We will demonstrate our approaches on three different path tracking controllers:
\begin{itemize}
    \item Stanley controller \cite{stanley}: 
    $$\delta = \varphi- \varphi_\text{ref}  + \arctan\left(\frac{k\cdot e}{v}\right)$$
    where $e$ is the shortest distance between vehicle front axle and the reference path: \mbox{$e=\min_{s}||\boldsymbol{p} - \boldsymbol{p}_\text{ref}(s)||$}, $\varphi_\text{ref}$ is the orientation of the reference path at the point $s$ of shortest distance and $k$ is a tuning parameter.
    \item Pure pursuit controller \cite{purepursuit}: 
    $$\delta = \arctan\left(\frac{2\cdot l\cdot e_\text{pp}}{l_\text{h}^2}\right)$$
    where $e_\text{pp}$ is the tracking error in a lookahead distance $l_\text{h}$ in front of the vehicles rear axle and $l$ is the vehicles wheelbase.
    \item Dubins robust control \cite{robopt}:
    $$\delta = \bar\delta \cdot \text{sign}(\sigma)$$
    where $\sigma(e, (\varphi - \varphi_\text{ref}), k_\text{rob})$ is a switching function implementing Dubins shortest path for reaching a reference path and $k_\text{rob}$ is a tuning parameter.
\end{itemize}

These controllers differ in their sensitivity to dead time. To demonstrate this, we simulate the controllers with the plant \eqref{eqn:kin_del}.
The vehicle is driving at a speed of $1$ m/s so that the kinetic effects are small and the control performance is only influenced by the dead time.
We test the controllers with dead time ($\Delta t_\text{in} + \Delta t_\text{out} = 0.4$ s) and without ($\Delta t_\text{in} = \Delta t_\text{out} = 0$ s). The controllers are simulated at a frequency of $100$ Hz ($\Delta t = 0.01$ s). The controllers are parameterized as:
\begin{align*}
k &= 3.0\text{s}^{-1}\\
l_\text{h} &= 1.45\text{m, } l=1.0\text{m}\\
k_\text{rob} &= 0.5
\end{align*}
Dubins robust control parameter $k_\text{rob}$ has been chosen in the centre of the permissible range. The Stanley control parameter $k$ has been chosen such that the reaching behavior without dead-time resembles the one of Dubins robust control, to increase comparability.  The lookahead distance $l_\text{h}$ has been chosen such that the region of convergence is large enough for the starting position.

\begin{figure}[tb] 
	\centering 
 \includegraphics[trim={0.7cm 1.8cm 0 0},scale=1]{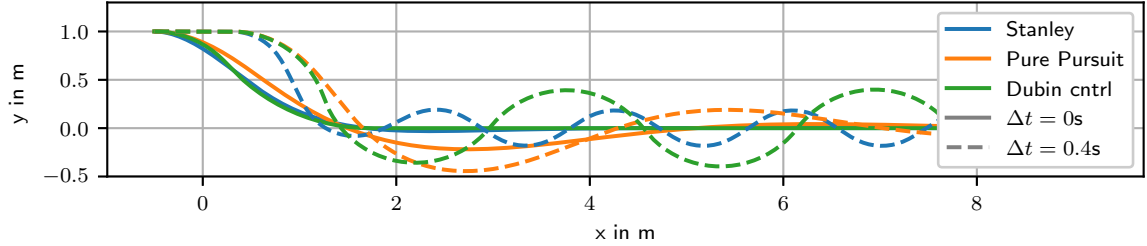}
	\caption{Paths of three different controllers approaching and tracking the reference path $\boldsymbol{p}_\text{ref} = \begin{bmatrix}x(s) & 0\end{bmatrix}^T$. The plant is simulated first without dead time and then with a dead time of $\Delta t = 0.4$ s.}
	\label{fig:demo_delay} 
\end{figure}

\begin{figure}[tb] 
	\centering 
 \includegraphics[trim={0.7cm 1.8cm 0 0},scale=1]{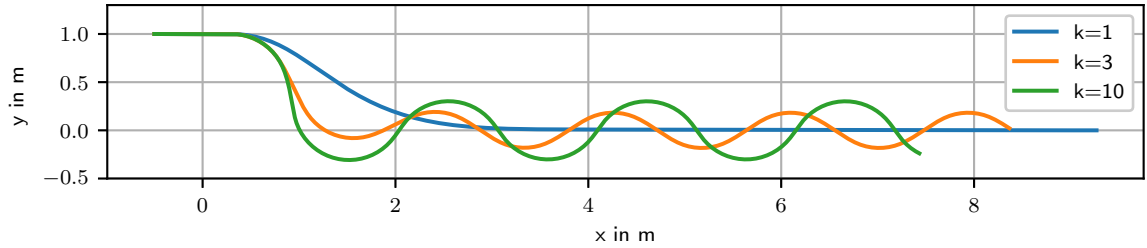}
	\caption{Paths of the Stanley controller approaching and tracking the reference path $\boldsymbol{p}_\text{ref} = \begin{bmatrix}x(s) & 0\end{bmatrix}^T$. The plant has a dead time of $\Delta t = 0.4$ s and the controller is parametrized with different values for $k$.}
	\label{fig:demo_delay_stan} 
\end{figure}

In Fig.~\ref{fig:demo_delay}, the paths produced by the different controllers are shown. In all 3 cases, the dead time leads to an oscillation around the reference path $\boldsymbol{p}_\text{ref} = \begin{bmatrix}x(s) & 0\end{bmatrix}^T$. The lookahead of the pure pursuit controller, acts as a dead-time reduction. However, a large lookahead leads to a slow reaching behavior. A similar behavior is also true for the other controllers: when they are tuned to reach the reference path faster, then the dead time has a greater impact on the tracking performance. To demonstrate this, we simulate the Stanley controller with different values of the parmeter $k$. The results are shown in Fig. \ref{fig:demo_delay_stan}

\subsubsection{Dead-time compensation}

As demonstrated, the controller typically can be tuned to more robustness to dead time in the plant system by decreasing the feedback gain or increasing a lookahead distance. However, this will decrease the reaching behavior and noise suppression. This robustness trade off is also discussed in \cite{robopt}.
Different to this approach, we want to compensate the dead time. With the use of a state prediction, it is possible to reduce the effect of dead time in the controlled loop without changing the controllers behavior. In short, we want to modify the plant such that it acts as if it had no dead time. In the following, we will give a mathematical formulation of this goal.

Mathematically, the plant model \eqref{eqn:kin_del} implements a mapping from the system state and the steering input to the output. We define:
\begin{align}
\boldsymbol{y}_{\text{del}} = f_P(\boldsymbol{x},\delta\cdot q^{-k_\text{I}})\cdot q^{-k_\text{O}}
\end{align}
When $v$ is constant, then $f_P$ is time invariant. In this case, the effect of input delay equals the effect of output delay. We define $k = k_\text{I} + k_\text{O}$:
\begin{equation}\label{eqn:timeinvariance}
    \boldsymbol{y}_\text{del} = f_P(x,\delta\cdot q^{-k_\text{I}})\cdot q^{-k_\text{O}} = f_P(x,\delta)\cdot q^{-k}
\end{equation}

The goal of the dead-time compensation $f_\text{dc}$ is to modify the feedback $\boldsymbol{y}_{\text{del}}$ such that the plant acts as if it had no delay:
\begin{equation}
    \boldsymbol{\hat y}_{} = f_{\text{dc}}(f_{P}(\boldsymbol{x},\delta)\cdot q^{-k}, \delta) = f_{\text{dc}}(\boldsymbol{y}_\text{del}, \delta)  \stackrel{!}{=} f_P(\boldsymbol{x},\delta)
\end{equation}

The structure of the control system with dead-time compensation is shown in Fig.~\ref{fig:cntrl_plant_dc}.
Due to the time invariance~\eqref{eqn:timeinvariance}, the dead-time compensation can be condensed to the task of predicting $\boldsymbol{y}$: 
\begin{equation}
    \hat{\boldsymbol{y}} = f_{\text{dc}}(\boldsymbol{y}_\text{del}, \delta) \stackrel{!}= \boldsymbol{y}_{\text{del}}\cdot q^{k}
\end{equation}
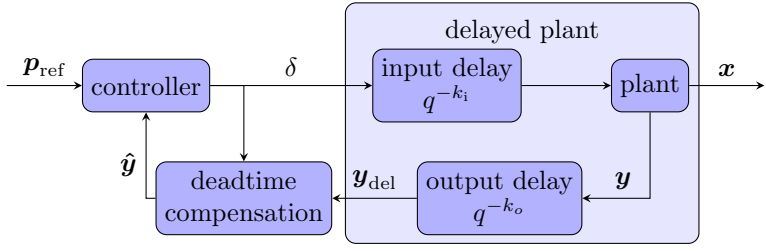
\begin{figure}[tb] 
	\centering 
	\footnotesize
\begin{tikzpicture}

\def\blockwidth{1.5cm}
\def\blockheight{0.7cm}

\node[draw, text depth=2.5cm,
rounded corners,
	minimum width = 4.7cm,
    minimum height=3.2cm,
    fill=blue!10,
align=center
] (pdel) at (5,-0.5){delayed plant};

\node[draw,
rounded corners,
	minimum width = \blockwidth,
    minimum height=\blockheight,
    fill=blue!30,
align=center
] (cntrl) at (0,0){controller};

\node[draw,
rounded corners,
	minimum width = \blockwidth*0.5,
    minimum height=\blockheight,
    fill=blue!30,
	align=center
] (p1) at ($(cntrl)+(4,0)$){input delay\\$q^{-k_\text{i}}$};

\node[draw,
rounded corners,
	minimum width = \blockwidth*0.5,
    minimum height=\blockheight,
    fill=blue!30,
	align=left
] (p2) at ($(p1)+(2.7,0)$){plant};

\node[draw,
rounded corners,
	minimum width = \blockwidth*0.5,
    minimum height=\blockheight,
    fill=blue!30,
	align=center
] (p3) at ($(p1)+(0.7,-1.5)$){output delay\\$q^{-k_o}$};

\node[draw,
rounded corners,
	minimum width = \blockwidth*0.5,
    minimum height=\blockheight,
    fill=blue!30,
	align=center
] (dc) at ($(p1)+(-2.7,-1.5)$){deadtime\\ compensation};




\draw[-stealth] (cntrl.east)--(p1.west) node[pos=0.5,above]{$\delta$};
\draw[-stealth] (p1.east)--(p2.west) node[pos=0.5,above]{};
\draw[-stealth](p2.east)--++(1,0)node[pos=0.5,above]{$\boldsymbol{x}$};
\draw[stealth-](p3.east)-|(p2.south)node[pos=0.3,above]{$\boldsymbol{y}$};
\draw[stealth-](dc.east)--(p3.west)node[pos=0.5,above]{$\boldsymbol{y}_\text{del}$};

\draw[stealth-](cntrl.west)--++(-1,0)node[pos=0.5,above]{$\boldsymbol{p}_\text{ref}$};

\draw[stealth-](cntrl.south)|-(dc.west)node[pos=0.3,left]{$\boldsymbol{\hat y}$};
\draw[stealth-](dc.north)|-(cntrl.east)node[pos=0.3,left]{};

\end{tikzpicture} 
	\caption{Structure of the controlled system with dead time in the plant and a dead-time compensation to modify the control feedback $\boldsymbol{y}_\text{del}$.}
	\label{fig:cntrl_plant_dc} 
\end{figure}

\subsection{Nonlinear motion prediction}

For predicting the undelayed plant output $\hat{\boldsymbol{y}}$ we use a plant model to compute future states. The plant model is a kinematic singletrack model \eqref{eqn:kin0}. 
When the controller is implemented in discrete time, the steering angle $\delta$ is constant during one timestep. The plant model resembles the discrete formulation of the plant \eqref{eqn:kin_del} without dead time, thus yielding the future state predictions $\boldsymbol{\hat y}\approx \boldsymbol{y}_\text{del}\cdot q^k$. With $\dot\psi = \tan(\delta)/l\cdot v$ we can write:
\begin{subequations}
    \label{eqn:pos_approx}
    \begin{align}
    \begin{split}
\boldsymbol{\hat p}_{i+1} - \boldsymbol{\hat p}_{i} &= \int_0^{\Delta t} \begin{bmatrix}
\cos(\hat\psi_{i} +\dot\psi_i\cdot \tau)\\ \sin(\hat\psi_{i} +\dot\psi_i\cdot \tau)
\end{bmatrix} \cdot v\cdot d\tau 
= \int_0^{\Delta t} \boldsymbol{R}(\hat\psi_{i})\cdot \begin{bmatrix}
\cos(\dot\psi_i\cdot \tau)\\ \sin(\dot\psi_i\cdot \tau)
\end{bmatrix} \cdot v\cdot d\tau \\
 &= \boldsymbol{R}(\hat\psi_{i})\cdot \begin{bmatrix}
    \sin(\dot\psi_i\cdot \Delta t) \\ (1-\cos(\dot\psi_i\cdot \Delta t))
\end{bmatrix}\cdot \frac{v}{\dot\psi_i}
\end{split}\\
\hat\psi_{i+1} &= \hat\psi_{i} + \dot\psi \cdot \Delta t \\
\boldsymbol{\hat y}_{i} &= \boldsymbol{\hat x}_{i} = \begin{bmatrix}
    \boldsymbol{\hat p}_{i} & \hat\psi_{i}
\end{bmatrix}^T
\end{align}
\end{subequations}


Where $\boldsymbol{R}$ is the rotation matrix:
\begin{equation}
    \boldsymbol{R}(\psi) = \begin{bmatrix}
        \cos(\psi) & -\sin(\psi) \\
        \sin(\psi) & \cos(\psi)
    \end{bmatrix}
\end{equation}

The plant model \eqref{eqn:pos_approx} is not internally stable (the state $\boldsymbol{\hat x}$ and output $\boldsymbol{\hat y}$ are ramp functions at steady state). In this case, a Smith predictor without modification is not applicable as $\boldsymbol{\hat y}$ will diverge \cite{smith}. There are solutions to regain internal stability as explained in \cite{hoelzl}. In this chapter, we will derive a prediction model that enables a suitable approximation of the future state despite divergence of $\boldsymbol{\hat y}$. This model is then modified to gain internal stability.

Independent of divergence of $\boldsymbol{\hat y}$, for small $k$\footnote{The error between the real vehicle dynamics and the stated kinematic approximation will accumulate over the steps $k$. Thus, for large $k$, the approximation will become inacurate.} we can state:

\begin{subequations}
\begin{align}
\boldsymbol{\hat p}_{i} - \boldsymbol{\hat p}_{i-k} &= \sum_{j=i}^{i+k}\boldsymbol{R}(\hat \psi_{j-k})\cdot \begin{bmatrix}
	\sin(\dot\psi_{j-k}\cdot\Delta t) \\ 1-\cos(\dot\psi_{j-k}\cdot\Delta t)
\end{bmatrix} \cdot \frac{v}{\dot\psi_{j-k}}\\
\boldsymbol{p}_{i+k} - \boldsymbol{p}_i &\doteq 
\sum_{j=i}^{i+k}\boldsymbol{R}(\psi_{j})\cdot \begin{bmatrix}
	\sin(\dot\psi_{j-k}\cdot\Delta t) \\ 1-\cos(\dot\psi_{j-k}\cdot\Delta t)
\end{bmatrix}\cdot \frac{v}{\dot\psi_{j-k}}\\
  \hat\psi_{i} - \hat\psi_{i-k} &= \sum_{j=i}^{i+k}\dot\psi_{j-k}\cdot \Delta t \doteq \psi_{i+k} - \psi_{i} \label{eqn:psi}
\end{align}
\end{subequations}
From \eqref{eqn:psi} follows for $j\in [i,i+k]$:
\begin{equation}
\psi_{i} - \hat\psi_{i-k} \doteq \psi_{j} - \hat\psi_{j-k} 
\end{equation}
Thus we can conclude:
\begin{equation}
\begin{split}
\boldsymbol{R}(\psi_i - \hat\psi_{i-k})\cdot\left(\boldsymbol{\hat p}_{i} - \boldsymbol{\hat p}_{i-k}\right) &\doteq \sum_{j=i}^{i+k}\boldsymbol{R}(\psi_{j} - \hat\psi_{j-k} + \hat \psi_{j-k})\cdot \begin{bmatrix}
	\sin(\dot\psi_{j-k}\cdot\Delta t) \\ 1-\cos(\dot\psi_{j-k}\cdot\Delta t)
\end{bmatrix}\cdot \frac{v}{\dot\psi_{j-k}} \\ &\doteq \boldsymbol{p}_{i+k} - \boldsymbol{p}_i,
\end{split}
\end{equation}
which leads to the state prediction:
\begin{equation}\label{eqn:statepred}
    \boldsymbol{x}_{i+k} \doteq \boldsymbol{x}_i + \begin{bmatrix}
        \boldsymbol{R}(\psi_{i} - \hat\psi_{i-k})\\ 1
    \end{bmatrix}^T \cdot (\boldsymbol{\hat x}_{i} - \boldsymbol{\hat x}_{i-k})
\end{equation}

\begin{figure}[tb] 
	\centering 
	\footnotesize
\begin{tikzpicture}

\def\blockwidth{1.5cm}
\def\blockheight{0.7cm}


\node[draw,
rounded corners,
	minimum width = \blockwidth,
    minimum height=\blockheight,
    fill=blue!30,
align=center
] (cntrl) at (0,0){controller};

\node[draw,
rounded corners,
	minimum width = \blockwidth*0.5,
    minimum height=\blockheight,
    fill=blue!30,
	align=center
] (p1) at ($(cntrl)+(4.5,0)$){delay\\$q^{-k}$};

\node[draw,
rounded corners,
	minimum width = \blockwidth*0.5,
    minimum height=\blockheight,
    fill=blue!30,
	align=left
] (p2) at ($(p1)+(2,0)$){plant};

\node[draw,
rounded corners,
	minimum width = \blockwidth*0.5,
    minimum height=\blockheight,
    fill=blue!30,
	align=center
] (p3) at ($(cntrl)+(3.5,-1.2)$){plant\\model};

\node[draw,
rounded corners,
	minimum width = \blockwidth*0.5,
    minimum height=\blockheight,
    fill=blue!30,
	align=left
] (p4) at ($(p3)+(-3.2,0)$){$\begin{bmatrix}\boldsymbol{R}(\varphi) \\ \boldsymbol{1}\end{bmatrix}^T \cdot \left(1-q^{-k}\right)$};


\node[] (add2) at ($(p4)+(-2.0,0)$){};
\filldraw[color=blue!30] (add2) circle (2.5pt);
\draw[color=black]  (add2) circle (2.5pt);

\draw[-stealth] (add2.north)|-($(cntrl.west)+(0,-0.1)$);
      node[pos=0.5,above]{};

\draw[-stealth] (cntrl.east)--(p1.west)
      node[pos=0.5,above]{$\delta_i$};
\draw[-stealth] (p1.east)--(p2.west)
      node[pos=0.5,above]{$\delta_{i-k}$};
\draw[-stealth](p2.east)--++(1,0)node[pos=0.5,above]{$\boldsymbol{x}_i$};
\draw[-stealth]($(cntrl.east)+(1,0)$)-|(p3.north)node[pos=0.5,above]{};
\draw[-stealth]($(p3.west)$)--(p4.east)node[pos=0.5,above]{$\boldsymbol{\hat{x}_{i}}$};
\draw[-stealth]($(p4.west)$)--(add2.east)node[pos=0.5,above]{};

\draw[-stealth]($(p2.east)+(0.5,0)$)|-++(-3,-2)-| (add2.south);
\draw[stealth-]($(cntrl.west)+(0,0.1)$)--++(-1,0)node[pos=0.5,above]{$\boldsymbol{p}_\text{ref}$};

\end{tikzpicture} 
	\caption{Structure of the control system with a delayed plant and dead-time compensation with the help of a prediction plant model.}
	\label{fig:deadtime_chart} 
\end{figure}
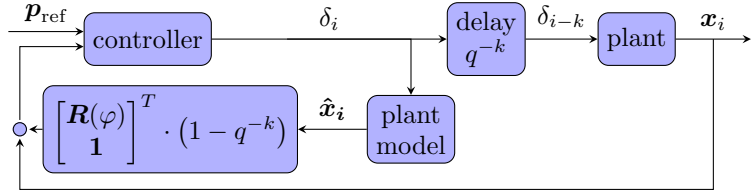

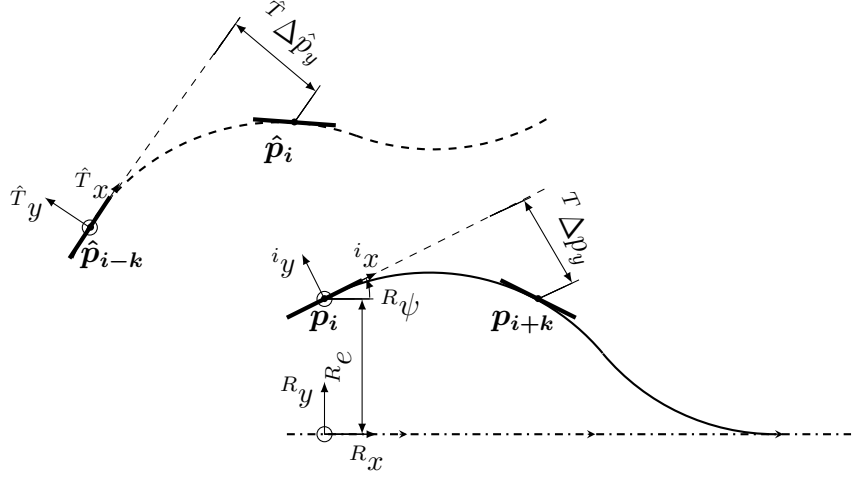
\begin{figure}[tb] 
	\centering 
	\begin{tikzpicture}[]

\def\lengthlabel[#1](#2)(#3)(#4:#5)(#6)
    {\draw (#2)--++(#4+90:#5+0.1);
     \draw (#3)--++(#4+90:#5+0.1);
    \draw (#2)--++(#4+90:#5-0.1);
     \draw (#3)--++(#4+90:#5-0.1);
    \draw (#2)--++(#4+90:#5)coordinate(l11);
     \draw (#3)--++(#4+90:#5)coordinate(l12);
    \draw [latex-latex](l11)--(l12)node[anchor=south, pos=0.5,rotate=#4]{#6};
 }

\def\centerarc[#1](#2)(#3:#4:#5)
    { \draw[#1] ($(#2)+({#5*cos(#3)},{#5*sin(#3)})$) arc (#3:#4:#5); }

\draw (-2.0,1.8) coordinate (v1) {};
\draw[ultra thick]($(v1)+(-0.5,-0.25)$) -- ($(v1)+(0.5,0.25)$) node[anchor=north, pos=0.5,yshift=0mm]{$\boldsymbol{p_i}$};
\draw ($(v1)-(0,0.02)$) node{\Huge$\cdot$};

\draw (0.84,1.8) coordinate (v2) {};
\draw[ultra thick]($(v2)+(-0.5,0.25)$) -- ($(v2)+(0.5,-0.25)$) node[anchor=north, pos=0.5,xshift=-2mm]{$\boldsymbol{p_{i+k}}$};
\draw ($(v2)-(0,0.02)$) node{\Huge$\cdot$};

\draw[thick,dash dot] (-2.5,0) --(5,0);
\draw[-stealth] (1.5,0)--++(0.1,0);
\draw[-stealth] (-1,0)--++(0.1,0);
\draw[-stealth] (4,0)--++(0.1,0);
\centerarc[thick](-0.6,-0.85)(40:90+27:3)
\centerarc[thick](4,3)(-90:-140:3)

\draw[](v1) -- ($(v1)+(0.6,0.3)$);
\draw (v1) -- ($(v1)+(0.65,0)$);
\draw[latex-] ($(v1)+(0.55,0.28)$) arc(22:0:0.7) node [anchor=west, pos=1,yshift=-1mm] {${}^R\psi$};
\lengthlabel[latex-latex](-2.0,0)(v1)(90:-0.5)(${}^R e$);

\lengthlabel[latex-latex](0.2,2.9)(v2)(-65:0.5)(${}^T \Delta p_y$);
\draw[dashed](v1) -- ($(v1)+(3,1.5)$);

\coordinate (origin) at (-2,0);
\draw [-latex] (origin) -- ++(0,.7) node[anchor=east,pos=0.8]{\small${}^R y$};
\draw [-latex] (origin) -- ++(0.7,0) node[anchor=north,pos=0.8]{\small${}^R x$};
\centerarc[](origin)(0:360:0.1);

\coordinate (origin) at (v1);
\draw [-latex] (origin) -- ++(-0.3,.6) node[anchor=east,pos=0.8]{\small${}^i y$};
\draw [-latex] (origin) -- ++(0.7,0.35) node[anchor=south,pos=0.8]{\small${}^i x$};
\centerarc[](origin)(0:360:0.1);

\def\dy{2};
\def\dx{-2};
\def\dpsi{30};
\centerarc[thick,dashed](-0.6+\dx,-0.85+\dy)(40+\dpsi:90+27+\dpsi+10:3)
\centerarc[thick,dashed](-0.55,6.79)(-90+\dpsi:-140+\dpsi:3)

\draw (-5.12,2.75) coordinate (e1) {};
\draw[ultra thick]($(e1)+(-0.27,-0.41)$) -- ($(e1)+(0.27,0.41)$) node[anchor=north, pos=0.5,xshift=3mm]{$\boldsymbol{\hat p_{i-k}}$};
\draw ($(e1)-(-0.01,0.02)$) node{\Huge$\cdot$};

\draw (-2.4,4.15) coordinate (e2) {};
\draw[ultra thick]($(e2)+(-0.55,0.04)$) -- ($(e2)+(0.55,-0.04)$) node[anchor=north, pos=0.5,xshift=-2mm]{$\boldsymbol{\hat p_{i}}$};
\draw ($(e2)-(0,0.02)$) node{\Huge$\cdot$};

\coordinate (origin) at (e1);
\draw [-latex] (origin) -- ++(-0.6,.4) node[anchor=east,pos=0.8]{\small${}^{\hat T} y$};
\draw [-latex] (origin) -- ++(0.4,0.6) node[anchor=east,pos=1]{\small${}^{\hat T} x$};
\centerarc[](origin)(0:360:0.1);

\draw[dashed] (e1)--++(2.,2.8);
\lengthlabel[latex-latex](-3.47,5.05)(e2)(-35:0.5)(${}^{\hat T} \Delta \hat p_y$);

\end{tikzpicture} 
	\caption{Concept of the presented predictor. The motion prediction $\Delta \boldsymbol{\hat p}$ is transformed from the predictor coordinate frame $\hat T$ to the vehicle coordinate frame $T$. The tracking error is measured in the reference coordinate frame $R$.}
	\label{fig:smith_coordinates} 
\end{figure}

This matches the structure of the Smith predictor except for the rotational transformation. The structure is also shown in Fig.~\ref{fig:deadtime_chart}.
However, keep in mind that the underlying assumptions are different to the Smith predictor and consequently the behavior is different. The Smith predictor requires a stable plant such that the plant model state $\boldsymbol{\hat x}$ will not diverge. In our approach, we apply the dead-time compensation on an unstable plant, resulting in a possible divergence of $\boldsymbol{\hat x}$. Due to the special structure of the plant model, we can transform the prediction $\Delta \boldsymbol{\hat x} = \boldsymbol{\hat x}_{i} - \boldsymbol{\hat x}_{i-k}$ to account for the difference $\boldsymbol{\hat x}_{i-k} - \boldsymbol{x}_i$. This concept is illustrated in Fig.~\ref{fig:smith_coordinates}. The position prediction $\Delta \boldsymbol{\hat p}$ is transformed from the prediction model coordinate frame $\hat T$ to the vehicle coordinate frame $T$.

\begin{figure}[tb] 
	\centering 
 \includegraphics[trim={0.7cm 0 0 0},scale=1]{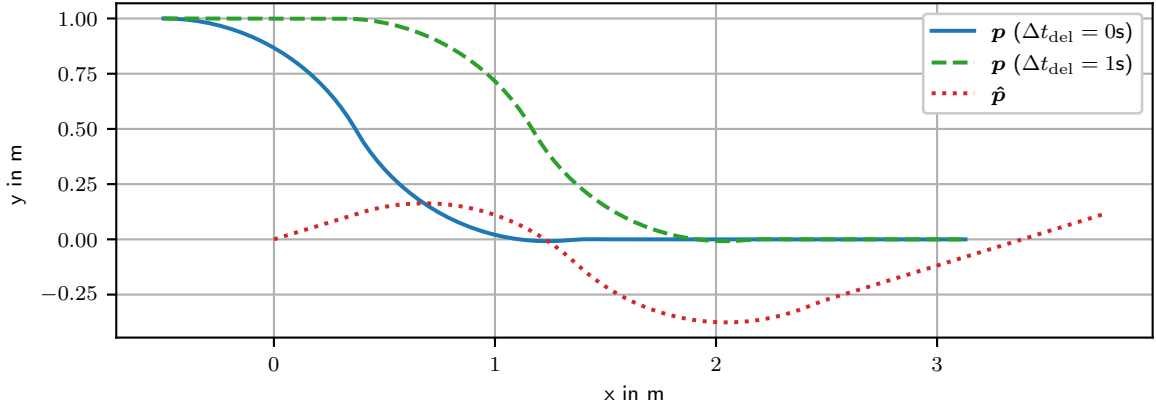}
	\caption{Paths of the Dubins robust controller with dead-time compensation. The simulation plant is a kinematic singletrack model first without dead time and then a dead time of $1$ s. $\boldsymbol{\hat p}$ is the path of the predictor plant for compensating the dead time}
	\label{fig:ideal_smith_p} 
\end{figure}

To make the behavior of the dead-time compensation more clear, we apply it on an ideal plant where the dead time is exactly known. The vehicle is steered to the reference path by the Dubins robust controller \cite{robopt}. As there are no unkown effects, the reaching path with compensated dead time $\boldsymbol{p}(\Delta t_\text{del}=0s)$ matches exactly the reaching path with dead time $\boldsymbol{p}(\Delta t_\text{del}=1s)$, except that the approaching sequence starts at a delay equal to the dead time. This is shown in Fig. \ref{fig:ideal_smith_p}. Additionally shown in this figure is the position of the plant model $\boldsymbol{\hat p}$ of the dead-time compensator. The initial model state is different to the real position and it does not converge to the real position at any time. However, the shape of the plant model path matches exactly the shape of the real path. Consequently, applying the rotation $\boldsymbol{R}(\psi_i - \hat\psi_{i-k})$, the prediction $\boldsymbol{\hat p}_{i} - \boldsymbol{\hat p}_{i-k}$ can be applied to the real position $\boldsymbol{p}_i$ to compute an (in this case) exact prediction of the real position $\boldsymbol{p}_{i+k}$.

\subsubsection{Implementation and stability}

The dead-time compensation requires implementation of the plant model~\eqref{eqn:pos_approx} and the state prediction~\eqref{eqn:statepred}. For the latter, the $k$ previous estimated states $\boldsymbol{\hat x}$ are stored in a queue \mbox{$\begin{bmatrix}
    \boldsymbol{\hat x}_{i-k} & \boldsymbol{\hat x}_{i-k+1} & \cdots & \boldsymbol{\hat x}_{i}
\end{bmatrix}^T$}. As discussed before, the state $\boldsymbol{\hat x}$ may diverge unboundedly. 
To prevent numeric difficulties, in each time step $i$, we shift the predicted path to the coordinate center $\boldsymbol{0}$. Furthermore, we project the orientation $\hat\psi$ onto the interval $[0,2\pi)$.
The prediction model state queue contains the $k$ last shifted position estimates and the $k$ last orientation estimates:
\begin{equation}
    \boldsymbol{\hat x}_{\text{q}} = \begin{bmatrix}
        \boldsymbol{\hat p_{\text{q},i,-k}} & \boldsymbol{\hat p_{\text{q},i,-k+1}} & \cdots & \boldsymbol{\hat p_{\text{q},i}} \\
        \hat \psi_{i-k} & \hat \psi_{i-k+1} & \cdots & \hat \psi_i
    \end{bmatrix}^T \in \mathbb{R}^{k\times 3}
\end{equation}
This queue describes the state increment from time step $i-k$ to each subsequent time step.
The prediction model state queue $\boldsymbol{\hat x}_{\text{q}}$ is computed as follows:
\begin{subequations}
\begin{align}
\boldsymbol{\hat x}_{\text{q},i+1} &= \begin{bmatrix}
    0 & 1 & 0 & \cdots &  & 0\\
    0 & 0 & 1 & 0 & \cdots & 0\\
      & & & \vdots\\
    0 & & & \cdots & & 1
\end{bmatrix}\cdot \boldsymbol{\hat x}_{\text{qs},i} + \begin{bmatrix}
    0 \\ \vdots \\ 0 \\ f_\text{pm}(\hat\psi_i, \delta_i)
\end{bmatrix}\\
f_\text{pm} &= \begin{bmatrix}
    \boldsymbol{R}(\hat\psi_{i}) \cdot \frac{v}{\dot\psi(\delta_i)} \cdot \begin{bmatrix}
    \sin(\dot\psi(\delta_i)\cdot \Delta t) \\ 1-\cos(\dot\psi(\delta_{i})\cdot \Delta t)
    \end{bmatrix} \\
    \dot\psi(\delta_i) \cdot \Delta t
    \end{bmatrix} \\
\boldsymbol{\hat p}_{\text{qs},i} &= \boldsymbol{\hat p}_{\text{q},i} - {\boldsymbol{\hat p}_{\text{q},i,-k}}\cdot \boldsymbol{1}_{2\times k}\label{eqn:shift} \\
\hat\psi_{\text{qs},i} &= \hat\psi_{i} \mod 2\pi
\end{align}
\end{subequations}
Where $f_\text{pm}$ is the state increment of the plant model in equation \eqref{eqn:pos_approx} with \mbox{$\boldsymbol{\hat x}_{i+1} = \boldsymbol{\hat x}_i + f_\text{pm}(\hat\psi_i, \delta_i)$} and $\boldsymbol{\hat x}_{\text{qs}}$ is the shifted state, containing positions $\boldsymbol{\hat p}_\text{qs}$ and orientations $\hat\psi_\text{qs}$. $\boldsymbol{1}_{k\times 1}$ is a $k\times 1$ matrix with value $1$ in every entry.

With \eqref{eqn:shift}, the predicted shifted position $\boldsymbol{\hat p}_{\text{qs},i,-k}$ is $\boldsymbol{0}$ by definition. The positions $\boldsymbol{\hat p}_{\text{q},i}$ will remain in a region around $\boldsymbol{0}$ that is bounded by the finite dead time $k$. Thus internal stability is obtained and the prediction model is BIBO stable.

The state prediction is computed according to \eqref{eqn:statepred} and using $\boldsymbol{\hat p}_{\text{qs},i,-k}=\boldsymbol{0}$:
\begin{equation}
    \boldsymbol{x}_{i+k} \doteq \boldsymbol{x}_i + \begin{bmatrix}
        \boldsymbol{R}(\psi_i - \hat\psi_{i-k}) \\ 1
    \end{bmatrix}^T \cdot \begin{bmatrix}
        \boldsymbol{\hat p}_{\text{qs},i}\\ \boldsymbol{\hat \psi}_{i} - \boldsymbol{\hat \psi}_{i-k}
    \end{bmatrix}
\end{equation}

\subsection{Evaluation}

To evaluate the effect of the dead-time compensation, we simulate a path tracking scenario with the mentioned controllers and the presented approach. The vehicle is modelled by a kinetic single track model which differs from the prediction model due to the tire slip. Additionally we induce errors in the dead-time and the vehicle parameters.

\subsubsection{Path tracking scenario}
\begin{figure}[tb] 
	\centering 
 \includegraphics[trim={0.7cm 0 0 0},scale=1]{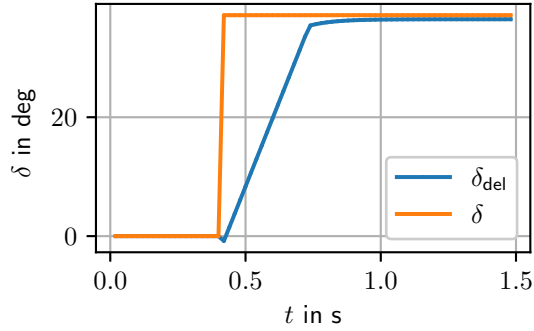}
	\caption{Step response of the steering model which is implemented as a third order low pass. The steering model is implemented with $\Delta t = 0.001$ s to resemble the steering system which is continuous in time.}
	\label{fig:steer_response} 
\end{figure}

The plant system is a kinetic single track model implementing a Pacejka tire model \cite{pacejka} and load transfer. It has a minimum turning radius of $3.8$ m.
The input and output delay is composed of a steering system and a constant dead time modelling delays due to computation time and data transmission. The steering system is modelled as a third order low pass filter with a step response as shown in Fig. \ref{fig:steer_response}. The constant dead time is set to a value of $0.27$ s. Note that the dynamics of the kinetic vehicle model also generate additional delay effects.

The reference path is composed of a right-hand-bend with radius of $27$ m and a left-hand-bend with an increasing radius of $27$ m to $100$ m and the vehicle is driving at a speed of $v=11.1$ m/s.

\subsubsection{Evaluation results}
\begin{figure}
\centering
\begin{subfigure}{0.8\textwidth}
    \includegraphics[trim={0.7cm 2cm 0 0},scale=1]{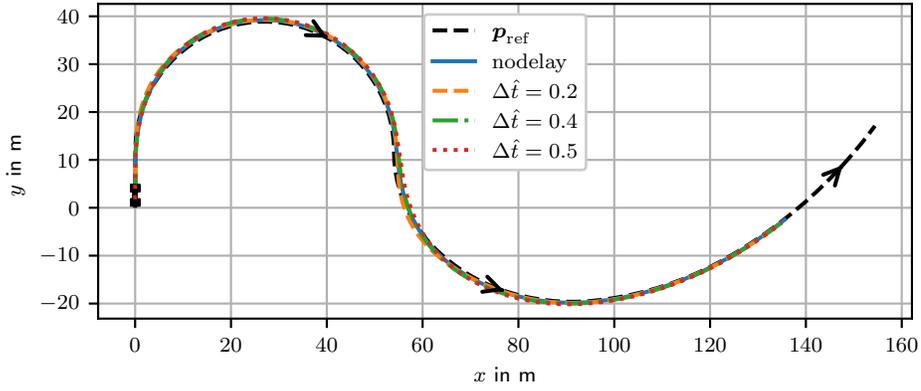}
    \caption{Paths driven by the vehicle.}
    \label{fig:pure_pursuit_p}
\end{subfigure}
\hfill
\begin{subfigure}{0.8\textwidth}
    \includegraphics[trim={0.7cm 0 0 0},scale=1]{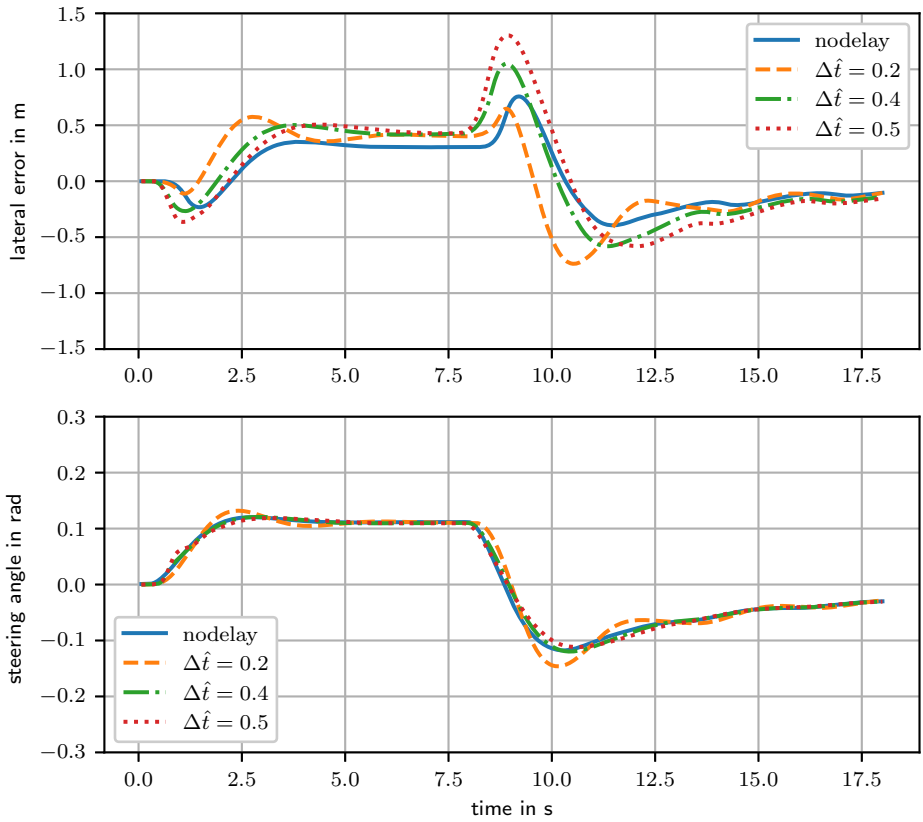}
    \caption{Tracking error and steering angle during the scenario.}
    \label{fig:pure_pursuit_e}
\end{subfigure}       
\caption{Simulation of the Pure pursuit controller. The plant model has first no dead time (blue) and then a dead time of $0.4$ s. The dead-time compensation is parameterized with different values for $\Delta \hat t$.}
\label{fig:pure_pursuit}
\end{figure}

\begin{figure}
\centering
\begin{subfigure}{0.7\textwidth}
    \includegraphics[trim={0.7cm 0.5cm 0 0cm},scale=1]{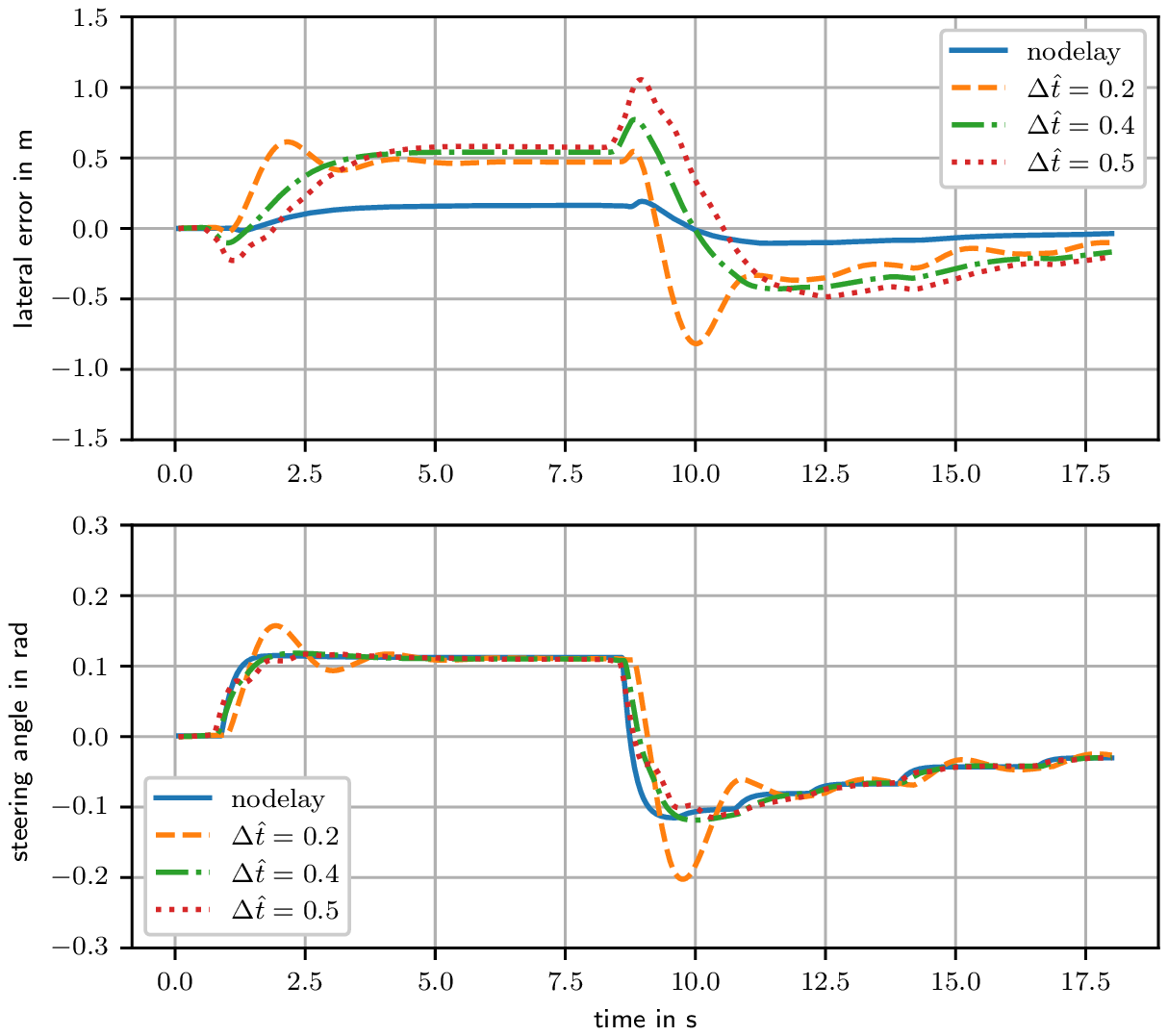}
    \caption{Stanley}
    \label{fig:stan}
\end{subfigure}
\hfill
\begin{subfigure}{0.7\textwidth}
    \includegraphics[trim={0.7cm 0.5cm 0 0},scale=1]{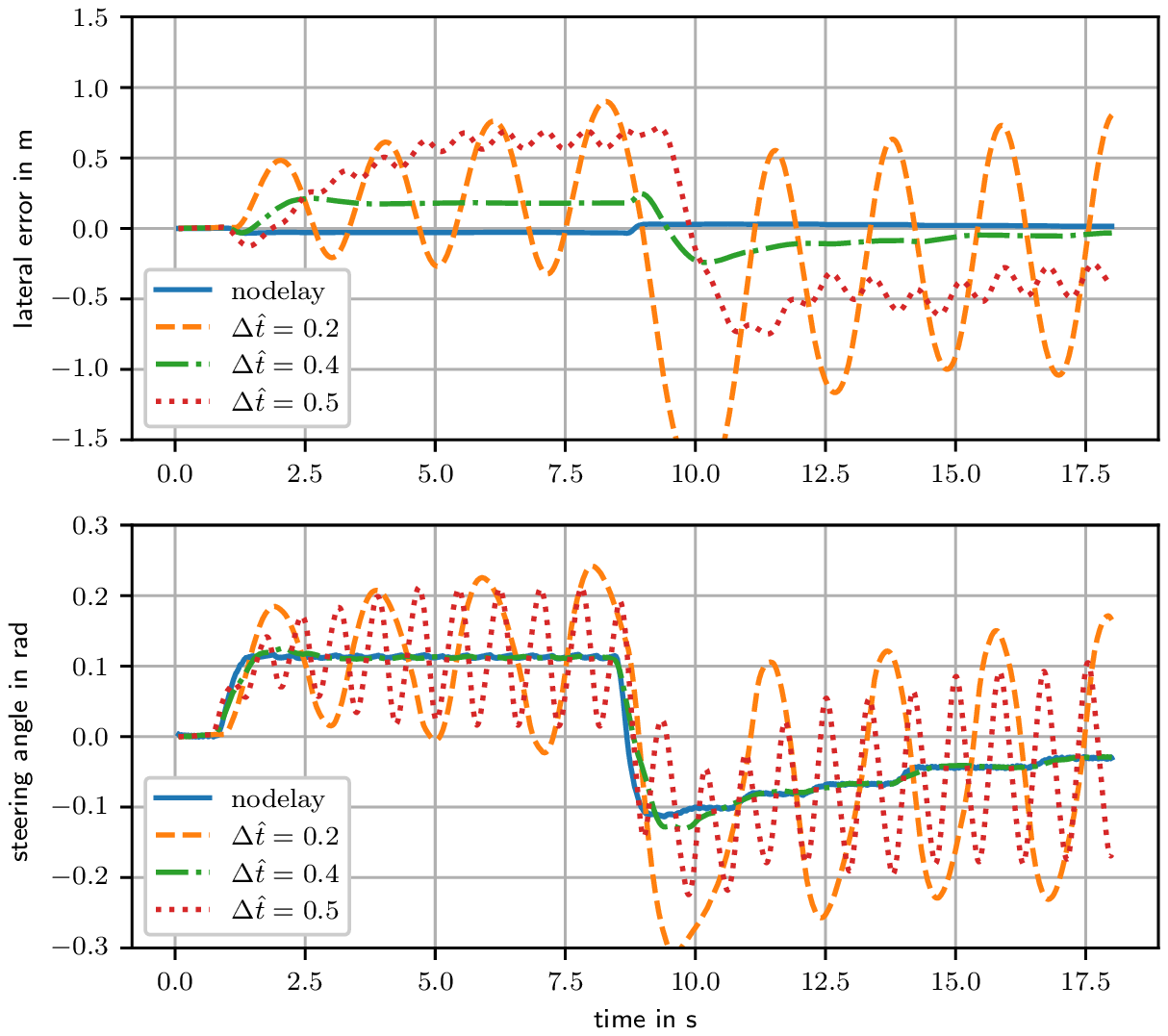}
    \caption{Dubins robust control}
    \label{fig:dubin}
\end{subfigure}       
\caption{Simulation of two controllers in the same scenario as in Fig. \ref{fig:pure_pursuit}.}
\label{fig:other_cntrl}
\end{figure}

We evaluate the path tracking scenario with the three controllers previously introduced. For each controller, the dead-time compensation is configured to compensate a dead time of
\begin{itemize}
    \item $\Delta t = 0.2$ s: The dead time is underestimated
    \item $\Delta t = 0.4$ s: Considering the low pass filter in combination with the steering rate, and the additional dead time in the plant system, this value approximately conforms with an effective dead time in the controlled system
    \item $\Delta t = 0.5$ s: The dead time is overestimated
\end{itemize}
The results are shown in Fig. \ref{fig:pure_pursuit} and Fig. \ref{fig:other_cntrl}. In all three controllers the effect of the dead-time compensation is qualitatively similar: the control performance with correctly parameterized dead-time compensation is worse than the control performance without dead time in the plant. However, the control performance with dead-time compensation is far better than without compensation and also better than with underestimation of the dead time. 

\begin{figure}
    \centering
    \includegraphics[trim={0.7cm 0 0 0},scale=1]{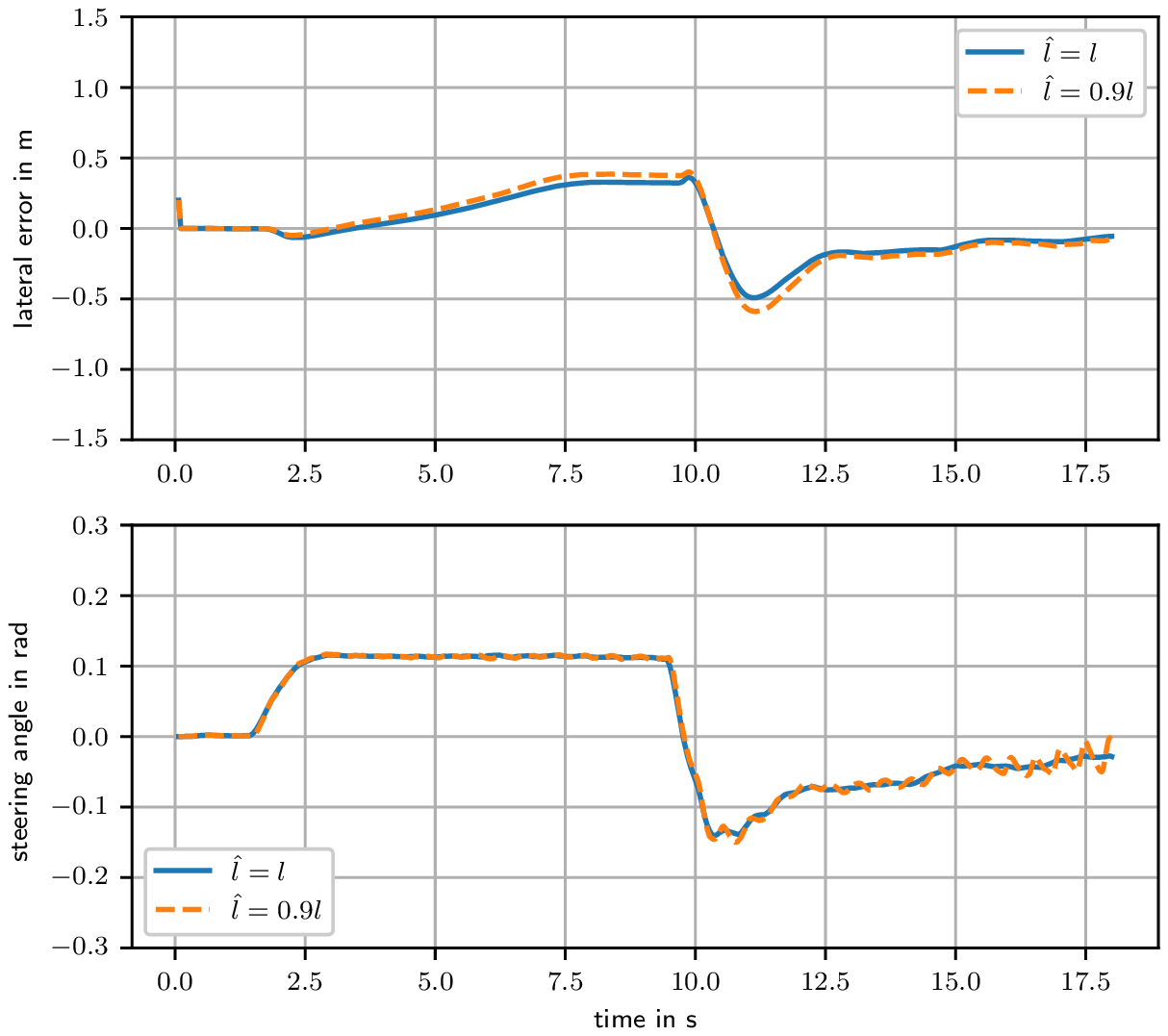}
    \caption{Simulation of the Dubins robust controller. The scenario is the same as in Fig. \ref{fig:pure_pursuit} except that we vary the wheelbase $\hat l$ in the predictor plant and the velocity $v$ is not constant.}
    \label{fig:eval_par}
\end{figure}

The dead-time compensation method heavily depends on a good approximation of the plant system. Therefore, we evaluate how the control performance changes when the vehicle parameters are not exactly known. More specifically, we set the vehicle wheelbase $\hat l$ in the predictor plant to
$$\hat l = 0.9 l$$
Moreover, in the derivation of the dead-time compensator we assumed constant velocity of the vehicle. To show the effect of a varying velocity, we add an acceleration from $5.5$ m/s to $12.5$ m/s at the beginning of the scenario.
The results are shown in Fig.~\ref{fig:eval_par}. The tracking error of the correctly parameterized controller is larger than in the previous simulation, due to the acceleration. This tracking error results from under-steering effects due to velocity dependent side slip.\\
The control error only slightly increases with the parameter uncertainty. When the wheelbase is underestimated, the predicted path has a larger curvature than the real path, resulting in a positive lateral error (with an overestimation of the wheelbase, the error would compensate the error caused by side slip, resulting in a smaller total error). Beside the change in tracking error, the deviation of the wheelbase estimate leads to oscillations in the steering angle at high velocity.

\subsection{Conclusion}
In this chapter, we addressed the significant challenge posed by dead time in the context of vehicle control, particularly in the application of classical path tracking controllers for autonomous vehicles. The simulation results consistently demonstrated the poor performance of various controllers when confronted with dead time, underscoring the pressing need for an effective compensation method.

We proposed a dead-time compensation method with a structure similar to a Smith predictor but with a state prediction that incorporates the specific nonlinearities of the kinematic vehicle model. This design makes a convergence of the prediction model not necessary and yet guarantees a bounded feedback correction. In simulation the method exhibited stability and significantly improved control performance also in presence of disturbances and parameter uncertainties.
The effective dead-time compensation method shows promise as a practical solution to reduce the effects of dead time in fast-paced environments where quick path planning and tracking is crucial.


\nocite*
\bibliographystyle{abbrv}
\bibliography{festl-chap3}

\begin{thebibliography}{1}

\bibitem{purepursuit}
C.~Coulter.
\newblock {Implementation of the pure pursuit path tracking algorithm}.
\newblock Technical report, Carnegie Mellon University, Pittsburgh, PA, 1992.

\bibitem{nonholonomic_outputdelay}
E.-H. Guechi, J.~Lauber, M.~Dambrine, G.~Klan{\v{c}}ar, S.~Bla{\v{z}}i{\v{c}}, E.-H. Guechi, J.~Lauber, {\textperiodcentered}.~M. Dambrine, M.~Dambrine, G.~Klan{\v{c}}ar, {\textperiodcentered}.~S. Bla{\v{z}}i{\v{c}}, and S.~Bla{\v{z}}i{\v{c}}.
\newblock {PDC Control Design for Non-holonomic Wheeled Mobile Robots with Delayed Outputs}.
\newblock {\em Intelligent Robot Systems}, 60:395--414, 2010.

\bibitem{hanus}
R.~Hanus, M.~Kinnaert, and J.~L. Henrotte.
\newblock {Conditioning technique, a general anti-windup and bumpless transfer method}.
\newblock {\em Automatica}, 23(6):729--739, 1987.

\bibitem{stanley}
G.~M. Hoffmann, C.~J. Tomlin, M.~Montemerlo, and S.~Thrun.
\newblock {Autonomous automobile trajectory tracking for off-road driving: Controller design, experimental validation and racing}.
\newblock In {\em Proceedings of the American Control Conference}, pages 2296--2301, 2007.

\bibitem{hoelzl}
S.~L. H{\"{o}}lzl.
\newblock {The case for the Smith-{\AA}str{\"{o}}m predictor}.
\newblock {\em Journal of Process Control}, 128:103026, 2023.

\bibitem{smith}
J.~E. Normey-Rico and E.~F. Camacho.
\newblock {\em {Control of dead-time processes}}.
\newblock Springer International Publishing, 2007.

\bibitem{smith_pred_cntrl}
J.~E. Normey-Rico, J.~{Goh Mez-Ortega}, and E.~F. Camacho.
\newblock {A Smith-predictor-based generalised predictive controller for mobile robot path-tracking}.
\newblock {\em Control Engineering Practice}, 7:729--740, 1999.

\bibitem{pacejka}
H.~B. Pacejka and E.~B. and.
\newblock The magic formula tyre model.
\newblock {\em Vehicle System Dynamics}, 21(sup001):1--18, 1992.

\bibitem{robopt}
K.~Tieber, J.~Rumetshofer, M.~Stolz, and D.~Watzenig.
\newblock {Sub-optimal and robust path tracking: a geometric approach}.
\newblock In {\em 2021 IEEE/RSJ International Conference on Intelligent Robots and Systems (IROS)}, pages 8381--8387, 2021.

\end{thebibliography}

\end{document}